\documentclass[conference, 10pt, twocolumn]{IEEEtran}

\usepackage{amsmath, amssymb}
\usepackage{graphicx}
\usepackage{booktabs}
\usepackage{cite}
\usepackage{hyperref}
\usepackage{xcolor}
\usepackage{multirow}
\usepackage{algorithm}
\usepackage{algorithmic}
\usepackage{tikz}
\usetikzlibrary{arrows.meta, positioning, calc}

\usepackage{lipsum}

\title{Evaluating the Temporal Detection Capability of Integrated Gradients
  Applied on Sound Classifier}

\author{
\IEEEauthorblockN{Martynas Dumpis}
\IEEEauthorblockA{
Department of Electronic Systems\\
Vilnius Gediminas Technical University\\
Vilnius, Lithuania\\
martynas.dumpis@vilniustech.lt}
\and
\IEEEauthorblockN{Tuomas Virtanen}
\IEEEauthorblockA{
Signal Processing Research Centre\\
Tampere University\\
Tampere, Finland\\
tuomas.virtanen@tuni.fi}
}

\begin{document}

\maketitle

\begin{abstract}
  Gradient-based attribution methods can highlight input regions important for
  neural network predictions, but their effectiveness for temporal sound event
  detection in audio classification has not been systematically
  evaluated. This paper assesses whether integrated gradients (IG) can
  temporally detect sound events when applied to a classifier trained without
  temporal supervision. We use synthetic polyphonic audio with ground truth
  timestamps to measure alignment between IG attributions and event
  boundaries. On a 10-class domestic sound dataset, IG achieves mean
  Intersection over Union (IoU) of 0.39, frame-level F1 of 0.52, and Pointing
  Game accuracy of 82.6\%.  For comparison, a framewise CNN trained with weak
  supervision (FW-WS, clip-level training labels) achieves 0.42 IoU, 0.55 F1,
  and 97.3\% PG, while a strongly supervised variant (FW-SS, frame-level
  training labels) reaches 0.45 IoU, 0.58 F1, and 97.9\% PG. Overall, these
  results suggest that post-hoc IG captures meaningful temporal activity
  patterns of sound events, with localization performance approaching models
  that explicitly produce frame-level predictions. All methods substantially
  outperform random and energy-based baselines.
\end{abstract}


\section{Introduction}
\label{sec:introduction}

Sound event detection (SED) aims to identify acoustic events in an audio
recording and estimate their temporal boundaries. In many applications,
obtaining precise onset and offset annotations is expensive, so detection
systems are often trained with weak supervision using only clip-level class
labels~\cite{turpault2019desed}. A common formulation treats weakly supervised
SED as multiple-instance learning: the model produces frame-level scores that
are aggregated over time to match the clip-level labels, enabling temporal
predictions without strong
annotations~\cite{wang2019pooling,deshmukh21_interspeech}.  In this context,
clip-level sound event classification provides the weak labels used to learn
detection. This work studies a more constrained setting: whether post-hoc
attribution from a classifier trained only for clip-level classification,
without temporal labels and without a framewise prediction head, can recover
information about temporal activities of sound events.

Gradient-based attribution techniques---integrated gradients (IG), layer-wise
relevance propagation (LRP), and gradient-weighted class activation mapping
(Grad-CAM)---can highlight input regions that influence model
predictions~\cite{akman2024audioxai,wang2024gradient}. Previous studies have
investigated explainability methods for speech and music signals: Becker et
al.~\cite{becker2024audiomnist} showed that LRP heatmaps align with phoneme
boundaries on AudioMNIST, and Seipel et al.~\cite{seipel2023lrp} applied LRP
for music instrument recognition. Frommholz et
al.~\cite{frommholz2023cbmi_xai} found attribution quality depends on input
representation, with waveform and spectrogram models highlighting different
features for identical predictions. Oguiza and
Parada-Cabaleiro~\cite{oguiza2024focal} proposed inherently interpretable
architectures as an alternative to post-hoc explanation. However, the existing
audio-XAI studies largely stop at explaining clip-level classification
decisions and leave unclear whether post-hoc attribution from a classifier can
serve as a reliable, temporally precise proxy for sound event activity without
temporal labels or a framewise prediction head.

This study evaluates whether post-hoc IG from a clip-level classifier can
recover temporal sound event activity in polyphonic audio. Using synthetic
soundscapes with ground-truth onset/offset annotations, we show that post-hoc
IG provides meaningful temporal detection, and we benchmark it against
framewise CNN baselines trained with weak supervision (FW-WS, clip-level
labels) and strong supervision (FW-SS, frame-level labels). Quantitative
evaluation of audio attributions is still limited, and existing benchmarks
largely focus on isolated sounds rather than overlapping
events~\cite{bolanos2025timelocalized}.

\section{Integrated gradients for sound event detection}
\label{sec:methodology}

\begin{figure}[t]
  \centering
  \includegraphics[width=0.8
  \columnwidth]{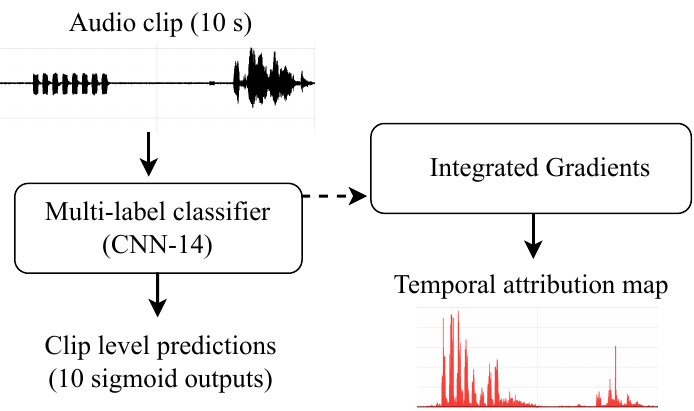}
  \caption{Block diagram of the proposed method. A 10~s audio clip is
    converted to a log-mel spectrogram and fed to a CNN14-based multi-label
    classifier to produce clip-level class scores. Integrated Gradients is
    then applied to the classifier to obtain a temporal attribution map.}
  \label{fig:pipeline}
\end{figure}

A clip-level multi-label classifier is trained using weak supervision, where
each 10~s audio clip is annotated only with a set of class labels indicating
which sound events are present somewhere in the clip, without onset/offset
timestamps. Given an input clip, the classifier outputs one sigmoid score per
class, representing the predicted presence probability of that class in the
clip. Integrated Gradients (IG) is then applied post-hoc to the trained
classifier to attribute each class score back to the input over time, yielding
a temporal importance map that can be thresholded into a frame-level detection
output (Fig.~\ref{fig:pipeline}).

 \subsection{Dataset}

 Synthetic polyphonic audio was generated using DESED~\cite{turpault2019desed}
 and Scaper~\cite{salamon2017scaper} library, providing precise temporal
 annotations for evaluation.. The dataset comprises 10-second audio clips
 sampled at 32~kHz, containing 1--3 foreground sound events, with events
 placed at random temporal positions such that overlap may occur when multiple
 events are present. The events belong to 10 domestic sound classes: Alarm
 bell ringing, Blender, Cat, Dishes, Dog, Electric shaver/toothbrush, Frying,
 Running water, Speech, and Vacuum cleaner. Source audio for foreground events
 was obtained from the DESED soundbank, while background sounds consisted of
 ambient domestic noise mixed at a reference level of $-55$~dB.

 Each clip contains 1--3 foreground events mixed with the background, where
 the number of events, their classes, and their temporal positions were
 sampled at random. Event durations follow the natural length of the source
 audio files (0.25--4.2~s). The signal-to-noise ratio (SNR) between each
 foreground event and the background noise was uniformly sampled between 15
 and 25~dB, ensuring clear audibility above the ambient noise floor.

 Training, validation and test sets were generated separately using distinct
 foreground audio pools from the DESED soundbank's training and evaluation
 splits, respectively, while sharing a common background audio pool. For the
 training set, 1000 audio clips were initially generated using the DESED
 soundscape generation pipeline, which subsequently removed files containing
 three simultaneous overlapping sound events, resulting in 823 training
 samples. For validation and test sets, 250 clips were generated and filtered
 using the same automated criterion, yielding 193 samples, which were split
 equally into validation (96 samples) and test (97 samples) subsets.

 \subsection{Model Architecture}

 The classification model combines a pretrained PANNs feature
 extractor~\cite{kong2020panns}, temporal aggregation, and a classification
 layer. Specifically, we use CNN14 pretrained on
 AudioSet~\cite{gemmeke2017audio}. Audio is converted to log-mel spectrograms
 using a 1024-sample Hamming window, 320-sample hop, and 64 mel bins spanning
 50~Hz to 14~kHz.  After the convolutional blocks, features are averaged
 across frequency and aggregated over time with global max pooling to obtain a
 2048-dimensional embedding.

 During training, CNN14 is frozen as a feature extractor. The original
 527-class output layer is replaced with a 10-class linear layer, trained with
 binary cross-entropy and sigmoid outputs for multi-label classification.

 \subsection{Integrated Gradients}

 IG~\cite{sundararajan2017axiomatic} is a gradient-based attribution method
 that assigns importance scores to input features by accumulating gradients
 along a path from a baseline input to the actual input. For an input $x$,
 baseline $x'$, and model output function $F$, the attribution for the $i$-th
 input dimension is defined as

 \begin{equation}
   \text{IG}_i(x) = (x_i - x'_i)\times \int_{\alpha=0}^{1}
   \frac{\partial F(x' + \alpha(x - x'))}{\partial x_i}\, d\alpha \text{.}
 \end{equation}

 The integral is approximated using $n$ discrete steps. In this work, IG
 attributions were computed using the Captum
 library~\cite{kokhlikyan2020captum} with $n = 50$ steps. The baseline input
 $x'$ was set to a zero-valued waveform representing silence, ensuring
 attributions reflect acoustic content relative to its absence.
 
 \begin{figure}[!b]
   \centering
   \includegraphics[width=0.9\columnwidth]{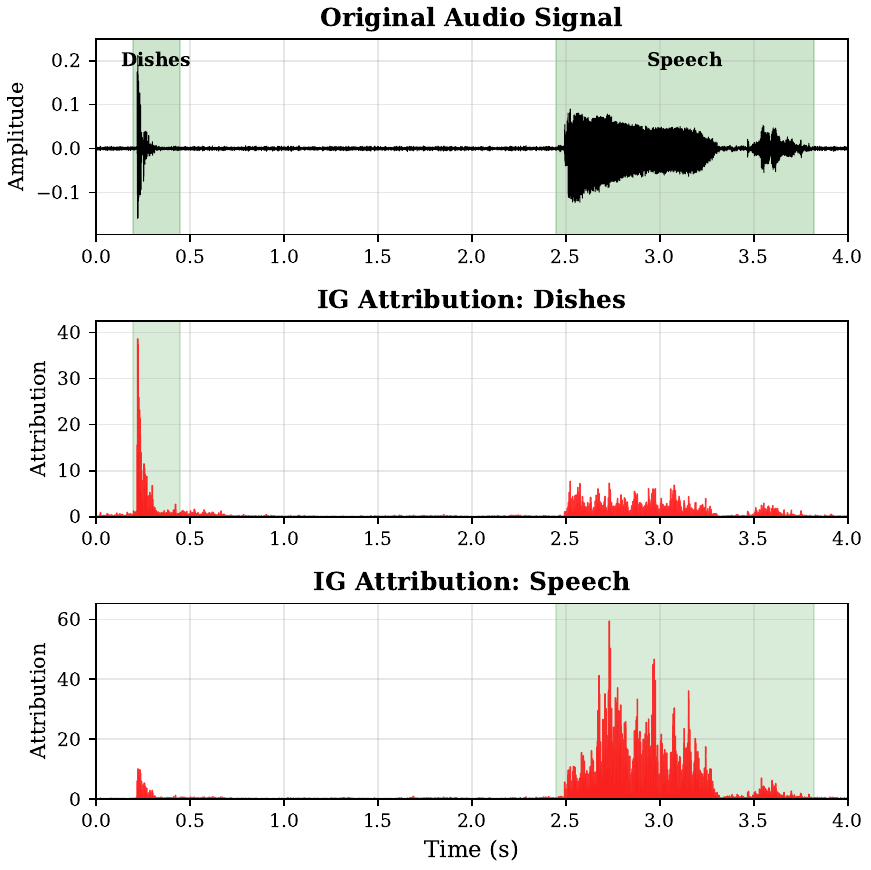}
   \vspace{-6pt}
   \caption{IG attribution magnitudes for a polyphonic test sample. Top:
     waveform with ground truth (green). Lower panels: class-specific
     importance magnitudes (red) showing temporal relevance for each predicted
     class. First 4 s shown.}
   \label{fig:ig_example}
 \end{figure}

 For multi-label classification, IG attributions were computed separately for
 each class with predicted probability exceeding 0.5. Given an input waveform
 of 320,000 samples, the resulting attribution vector has identical
 dimensionality, with each sample assigned a signed importance score. Positive
 values indicate features increasing class probability; negative values
 indicate suppressive contributions. Absolute values were used to obtain
 importance magnitudes for subsequent analysis. To illustrate the resulting
 attributions, Fig.~\ref{fig:ig_example} shows class-specific IG scores for a
 representative polyphonic test clip. The attributions concentrate within the
 ground-truth event boundaries, with sustained events producing broader
 high-importance regions and transient events producing sharper peaks.

\section{Experiments}
\label{sec:experiments}
 \subsection{Framewise Temporal Baselines}

 To provide a learned temporal detection baseline for comparison with IG
 attributions, a framewise variant of CNN14 is implemented. This follows the
 neural MIL formulation commonly used in weakly supervised SED, where
 frame-level event scores are aggregated over time (often by max pooling) to
 produce clip-level
 predictions~\cite{wang2019pooling,deshmukh21_interspeech}. The standard CNN14
 architecture pools over time before classification, producing a single
 clip-level prediction. The framewise variant removes temporal pooling before
 classification and applies the classifier independently at each time frame,
 producing predictions of shape $(B, T, C)$; clip-level scores are obtained
 via temporal max pooling after classification.

 We train two framewise variants. The weakly supervised model (FW-WS) is
 trained using clip-level labels, where frame-level predictions are aggregated
 by temporal max pooling to match clip-level targets. The strongly supervised
 model (FW-SS) is trained using frame-level event activity labels at the
 same temporal resolution.

 For 10-second audio clips at 32~kHz with the specified mel-spectrogram
 parameters, the architecture outputs predictions at 31 temporal frames,
 corresponding to approximately 0.32-second resolution. During both training
 and evaluation, the clip-level prediction for each class $c$ is computed as:

 \begin{equation}
   p_{\text{clip}}(c) = \max_{t=1}^{T} p_t(c)
 \end{equation}

 \noindent where $p_t(c)$ denotes the predicted probability for class $c$ at
 frame $t$.

 The framewise model uses the same transfer learning setup: CNN14 weights
 pretrained on AudioSet are loaded, the base convolutional layers are frozen,
 and the 10-class classification layer is trained with binary
 cross-entropy. This architecture provides explicit temporal predictions
 without requiring post-hoc attribution methods, serving as a baseline for
 evaluating whether IG attributions capture meaningful temporal information
 comparable to directly learned detection.

 \subsection{Evaluation Metrics}

 We report multi-label classification performance and evaluate temporal
 detection quality of IG attributions.

 \textit{1) Classification sanity check:} Multi-label classification
 performance is assessed using per-class precision, recall, and F1-score. A
 prediction threshold of 0.5 is applied to sigmoid outputs for binary
 predictions.

 \textit{2) Temporal Detection Metrics:} Temporal detection quality is
 evaluated using four metrics: IoU, frame-level F1-score, and the Pointing
 Game metric.  All metrics are computed at 100 ms frame resolution, standard
 for SED evaluation~\cite{mesaros2016metrics_polyphonic}.

 IoU measures alignment between attributions and ground truth events:

 \begin{equation}
   \text{IoU} = \frac{|M_{\text{attr}} \cap M_{\text{GT}}|}{|M_{\text{attr}} \cup M_{\text{GT}}|}
 \end{equation}

 \noindent where $M_{\text{attr}}$ is the binary mask of attribution values
 above a percentile threshold $\tau$, and $M_{\text{GT}}$ is the ground truth
 event activity.  Attributions are aggregated to frame-level by averaging
 absolute values within each 100 ms window before thresholding. The optimal
 threshold was selected on the validation set through sensitivity analysis
 across percentiles 1--99.

 Frame-level F1-score treats temporal detection as binary classification at
 each frame. Precision measures what fraction of frames marked as important by
 the attribution actually correspond to sound events, while recall measures
 what fraction of actual event frames are captured:

 \begin{equation}
   \text{Precision} = \frac{\text{TP}}{\text{TP} + \text{FP}}, \quad
   \text{Recall} = \frac{\text{TP}}{\text{TP} + \text{FN}}
 \end{equation}

 \noindent where TP, FP, and FN are computed by comparing $M_{\text{attr}}$
 against $M_{\text{GT}}$ at each frame.

 Pointing Game accuracy determines whether the maximum attribution falls
 within ground truth boundaries:

 \begin{equation}
   \text{PG} = \mathbf{1}[t_{\max} \in \text{GT events}]
 \end{equation}

 \noindent where $t_{\max} = \arg\max_t |\text{attr}(t)|$.

 IG detection is compared against two baselines: random attribution with
 uniformly sampled scores, and energy-based attribution using signal
 amplitude. Framewise CNN baselines are trained under weak and strong
 supervision as reference detectors.

 \subsection{Evaluation Protocol}

 IG attributions were computed for all test samples. For each sample, the
 clip-level classifier was first applied, and attributions were generated only
 for classes with predicted probability exceeding 0.5. Classes below this
 threshold were considered undetected and excluded from attribution
 analysis. Framewise models (WS/SS) were evaluated on the same test set to
 produce frame-level predictions for each detected class instance.

 To determine optimal binarization thresholds for IoU and F1 computation,
 percentile thresholds from 1 to 99 were evaluated on the validation
 set. Metrics were computed across all test samples and averaged for each
 threshold value.

 Detection performance was analyzed per sound class. IoU, F1-score, and
 Pointing Game metrics were computed separately for each of the 10 classes,
 averaged over test samples containing that class, and then macro-averaged
 over classes to obtain overall scores.

 Models were trained in PyTorch for 100 epochs with early stopping (patience
 10) using Adam ($10^{-3}$) and batch size 16; CNN14 base layers were frozen
 and only the classification layers were trained.

 \section{Results}
 \label{sec:results}
 \subsection{Classification Performance}

 Table~\ref{tab:classification} reports test-set classification performance.
 Speech achieves the highest F1 (0.95), followed by Cat (0.90), while Blender
 (0.27) and Frying (0.29) have low recall.

  \begin{table}[!b]
   \centering
   \caption{Multi-label classification performance}
   \label{tab:classification}
   \begin{tabular}{lccc}
     \toprule
     \textbf{Class} & \textbf{Precision} & \textbf{Recall} & \textbf{F1} \\
     \midrule
     Alarm bell ringing & 1.00 & 0.75 & 0.86 \\
     Blender & 1.00 & 0.15 & 0.27 \\
     Cat & 1.00 & 0.81 & 0.90 \\
     Dishes & 1.00 & 0.57 & 0.72 \\
     Dog & 1.00 & 0.56 & 0.71 \\
     Electric shaver & 0.91 & 0.77 & 0.83 \\
     Frying & 1.00 & 0.17 & 0.29 \\
     Running water & 1.00 & 0.46 & 0.62 \\
     Speech & 0.92 & 0.98 & 0.95 \\
     Vacuum cleaner & 0.86 & 0.46 & 0.60 \\
     \bottomrule
   \end{tabular}
 \end{table}

 The performance variation across classes aligns with known challenges in
 domestic sound event detection~\cite{serizel2020desed}. Blender and Frying
 both produce broadband noise that overlaps spectrally, making them difficult
 to distinguish. In contrast, Speech achieves the highest F1-score due to its
 distinctive harmonic structure, which CNN features capture
 effectively~\cite{kong2020panns}.

 \subsection{Temporal Detection Comparison}

 Table~\ref{tab:localization} compares temporal detection methods. IG achieves
 mean IoU of 0.39, F1 of 0.52, and Pointing Game (PG) accuracy of 82.6\%. The
 framewise CNN baselines achieve higher scores, with the WS model reaching
 0.42 IoU, 0.55 F1 (97.3\% PG) and the SS model 0.45 IoU, 0.58 F1 (97.9\% PG),
 serving as an upper bound trained with frame-level labels. All methods
 substantially outperform the random and energy baselines.

 The small IoU gap between IG and the WS baseline suggests that post-hoc
 attribution can recover temporally informative regions without explicit
 temporal supervision. This extends prior audio explainability
 work~\cite{haunschmid2020audiolime,paissan2024listenable} toward quantitative
 temporal detection. The main difference is PG accuracy, indicating that IG is
 less consistent in placing peak attribution inside ground-truth events, even
 when overall overlap is comparable.

 \subsection{Per-Class Detection}

 \begin{table}[t]
   \centering
   \caption{Temporal detection performance}
   \label{tab:localization}
   \begin{tabular}{lcccc}
     \toprule
     \textbf{Method} & \textbf{Mean IoU} & \textbf{F1} & \textbf{Std IoU} & \textbf{PG} \\
     \midrule
     IG & 0.39 & 0.52 & 0.23 & 82.6\% \\
     FW-WS & 0.42 & 0.55 & 0.24 & 97.3\% \\
     FW-SS & 0.45 & 0.58 & 0.24 & 97.9\% \\
     \midrule
     Random baseline & 0.19 & 0.30 & 0.11 & 28.3\% \\
     Energy baseline & 0.16 & 0.24 & 0.16 & 15.9\% \\
     \bottomrule
   \end{tabular}
 \end{table}
 Table~\ref{tab:perclass_iou} compares per-class IoU and F1 for IG and the
 framewise baselines (WS/SS). For IG, Electric shaver/toothbrush achieves the
 highest IoU (0.67) and F1 (0.79), followed by Blender (0.63 IoU, 0.75 F1),
 while Dishes (0.20 IoU, 0.31 F1) and Speech (0.32 IoU, 0.46 F1) show the
 lowest localization performance. The framewise baselines improve both IoU and
 F1 for several classes, most notably Speech (IoU: 0.41 WS, 0.43 SS; F1: 0.55
 WS, 0.57 SS), whereas Dishes remains challenging for all methods.

 \begin{table}[!b]
   \centering
   \caption{Per-class temporal detection}
   \label{tab:perclass_iou}
   \begin{tabular}{lcccccc}
     \toprule
     \multirow{2}{*}{\textbf{Class}} &
                                       \multicolumn{2}{c}{\textbf{IG}} &
                                                                         \multicolumn{2}{c}{\textbf{FW-WS}} &
                                                                                                              \multicolumn{2}{c}{\textbf{FW-SS}} \\
     \cmidrule(lr){2-3}\cmidrule(lr){4-5}\cmidrule(lr){6-7}
                                     & \textbf{IoU} & \textbf{F1} & \textbf{IoU} & \textbf{F1} & \textbf{IoU} & \textbf{F1} \\
     \midrule
     Alarm bell ringing & 0.44 & 0.57 & 0.45 & 0.58 & 0.46 & 0.59 \\
     Blender & 0.63 & 0.75 & 0.67 & 0.78 & 0.69 & 0.82 \\
     Cat & 0.47 & 0.61 & 0.54 & 0.67 & 0.55 & 0.68 \\
     Dishes & 0.20 & 0.31 & 0.20 & 0.31 & 0.24 & 0.36 \\
     Dog & 0.45 & 0.57 & 0.32 & 0.45 & 0.34 & 0.46 \\
     Electric shaver & 0.67 & 0.79 & 0.66 & 0.77 & 0.66 & 0.77 \\
     Frying & 0.40 & 0.57 & 0.52 & 0.68 & 0.52 & 0.68 \\
     Running water & 0.49 & 0.62 & 0.45 & 0.58 & 0.49 & 0.62 \\
     Speech & 0.32 & 0.46 & 0.41 & 0.55 & 0.43 & 0.57 \\
     Vacuum cleaner & 0.51 & 0.65 & 0.45 & 0.60 & 0.52 & 0.65 \\
     \bottomrule
   \end{tabular}
 \end{table}

 Continuous sounds (e.g., Blender, Electric shaver/toothbrush, Running water)
 achieve higher IoU and F1 than more variable events (Speech, Dishes). This
 behavior may also reflect the global max pooling used by the classifier: the
 model is encouraged to retain the most discriminative temporal evidence for
 clip-level recognition, so IG can emphasize salient peaks rather than the
 full event extent. Notably, Speech attains high clip-level classification
 performance (Table~\ref{tab:classification}) but lower localization scores
 (IoU 0.32, F1 0.46 for IG), suggesting that detecting presence is easier than
 estimating temporal extent in polyphonic mixtures.

 \subsection{Threshold Sensitivity}

\begin{figure}[t]
   \centering
   \includegraphics[width=0.98\columnwidth]{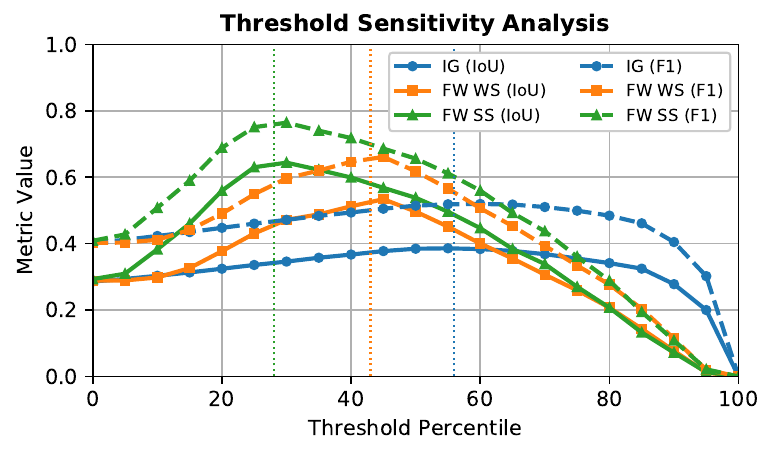}
   \vspace{-10pt}
   \caption{Threshold sensitivity of temporal detection. IoU (solid) and
     frame-level F1 (dashed) versus percentile threshold $\tau$ for IG and the
     framewise baselines trained with weak (FW-WS) and strong (FW-SS)
     supervision.}
   \label{fig:threshold}
 \end{figure}

 Figure~\ref{fig:threshold} shows IoU (solid) and frame-level F1 (dashed) as a
 function of percentile threshold $\tau$ for IG and the framewise baselines.
 For IG, peak IoU occurs at the 56th percentile (0.39) and peak F1 at the 57th
 percentile (0.52); the commonly used 80th percentile yields lower values (IoU
 0.34). In contrast, FW-WS peaks at the 43rd percentile (IoU 0.54, F1 0.67),
 while FW-SS peaks at the 28th--29th percentiles (IoU 0.65 at 28th, F1 0.77 at
 29th), indicating that the optimal operating point depends on the detector
 and metric.

 The substantial gap between optimal and default thresholds has practical
 implications for XAI evaluation. In weakly supervised detection for images,
 Zhou et al.~\cite{zhou2016cam} used a fixed 20\% threshold of maximum
 activation, while Selvaraju et al.~\cite{selvaraju2017gradcam} applied
 similar heuristics for Grad-CAM visualization. Our results suggest that such
 fixed thresholds may be suboptimal for audio temporal detection, where event
 boundaries are less sharply defined than object edges in images. The optimal
 threshold likely depends on the acoustic characteristics of target events and
 the temporal resolution of the underlying model. A recent survey on
 gradient-based attribution~\cite{wang2024gradient} notes that threshold
 selection remains an open challenge, with most evaluation protocols relying
 on dataset-specific tuning.

 \section{Conclusion}
 \label{sec:conclusion}

 We evaluated IG as a post-hoc method for temporal sound event detection from
 clip-level sound event classification. The main finding is that IG produces
 attribution maps that reveal temporal activity of sound events despite being
 computed from a classifier without temporal supervision, achieving 0.39 mean
 IoU and 0.52 F1. This approaches a framewise CNN trained with weak
 supervision (FW-WS: 0.42 IoU, 0.55 F1), while a strongly supervised framewise
 model provides an upper bound (FW-SS: 0.45 IoU, 0.58 F1). Both IG and the
 framewise baselines substantially outperform the random and energy baselines
 (0.19/0.16 IoU; 0.30/0.24 F1), indicating that the clip-level classifier
 retains information about when events occur, even though it is trained only
 to predict which events are present.

 Analysis across sound classes showed that stationary sounds such as Blender
 and Running water are detected more reliably than transient events like
 Speech or Dishes. We also found that threshold selection has substantial
 impact on metrics---the optimal 56th percentile outperformed the commonly
 used 80th percentile by about 13\% relative IoU (0.39 vs 0.34).

 Future work could compare IG against other attribution methods such as
 Grad-CAM, LRP, or perturbation-based approaches, as well as attention-based
 approaches~\cite{miyazaki2020selfattention}. Evaluation on real recorded and
 lower-SNR mixtures is also needed, since the present study uses synthetic
 soundscapes with relatively clear foreground events.

\section*{Acknowledgments}

This research was co-funded by the European Union under the Horizon Europe
programme, grant agreement No. 101059903, and by European Union funds for the
period 2021–2027 and the state budget of the Republic of Lithuania under
financial agreement No. 10-042-P-0001.
\bibliographystyle{IEEEtran}

\bibliography{IG_audio_detector}

\end{document}